\documentclass[a4paper,11pt]{article}
\pdfoutput=1 % if your are submitting a pdflatex (i.e. if you have
\usepackage{jheppub} % for details on the use of the package, please
\usepackage[T1]{fontenc} % if needed
\usepackage{url}
\usepackage{comment}
\usepackage{subfig}
\usepackage{tabularx}
\usepackage{graphicx}

\title{\boldmath Particle Track Reconstruction using Geometric Deep Learning}

\author[a,1]{Yogesh Verma,\note{Corresponding author.}}
\author[a,2]{Satyajit Jena,\note{Corresponding author.}}
\affiliation[a]{Department of Physical Science, Indian Institute of Science Education and Research (IISER)\\ Sector-81, Knowledge City, Mohali, India }

\emailAdd{ms16027@iisermohali.ac.in}
\emailAdd{sjena@iisermohali.ac.in}

\abstract{Muons are the most abundant charged particles arriving at sea level originating from the decay of secondary charged pions and kaons. These secondary particles are created when high-energy cosmic rays hit the atmosphere interacting with air nuclei initiating cascades of secondary particles which led to the formation of extensive air showers (EAS). They carry essential information about the extra-terrestrial events and are characterized by large flux and varying angular distribution.  To account for open questions and the origin of cosmic rays, one needs to study various components of cosmic rays with energy and arriving direction. Because of the close relation between muon and neutrino production, it is the most important particle to keep track of. We propose a novel tracking algorithm based on the Geometric Deep Learning approach using graphical structure to incorporate domain knowledge to track cosmic ray muons in our 3-D scintillator detector. The detector is modeled using the GEANT4 simulation package and EAS is simulated using CORSIKA (COsmic Ray SImulations for KAscade) with a focus on muons originating from EAS. We shed some light on the performance, robustness towards noise and double hits, limitations, and application of the proposed algorithm in tracking applications with the possibility to generalize to other detectors for astrophysical and collider experiments.

}

\begin{document} 
\maketitle
\section{Introduction}
\label{sec:intro}

Particle physics experiments are designed to understand the fundamental laws of nature by observing elementary particles, probing the interactions of elementary particles in vast quantities of particle collision data. The data produced by the experiments contains about O(10k) particles which leave O($10^{2}$k) hits and interactions in the detectors. Being at the frontiers of high energy physics, experiments must increasingly search for regimes of higher energy, higher data volume, and higher data density. One major component of the data analysis is the reconstruction of particle trajectories in tracking detectors. Tracking algorithms must be able to identify as many of these trajectories as possible while prioritizing the particles coming from the high energy interactions having large transverse momentum.

Cosmic muons are highly penetrating particles, which are created by the interaction of high energy cosmic particles on the upper atmosphere of Earth. Muons originate from the decay of secondary charged pions and kaons, which make up the majority of the charged particles reaching sea level. To study the cosmic rays we need to study their muon and electron components carefully in detectors by accurately tracking their interactions, path, etc. The close relationship between muon and neutrino production, it is the most important particle to keep track of in the detector and finding the vertex of decay particles. Experiments like IceCube~\cite{icecube}, GRAPES-3~\cite{grapes}, ARGO-YBJ~\cite{argo} are designed to study neutrino and cosmic rays with muon detector for muon tracking. Moreover, new detector proposals like MATHUSLA~\cite{mathusla}, INO ICAL~\cite{Kumar_2017} are aimed to study long-lived particles and make important contributions to cosmic ray physics. 

With the advent of High-Luminosity experiments, due to the explosion of hit combinatorics track reconstruction presents a challenging pattern recognition problem. The track reconstruction algorithms are inherently serial and highly time-consuming.  It is thus worthwhile to investigate fast algorithms for track reconstruction by new methods based on Deep Learning~\cite{lecun2015deep}. Motivated by the high computational cost, we investigate novel machine learning solutions in particle tracking by applying them to cosmic ray muon track reconstruction in a scintillator detector. Computer Vision techniques like CNNs are applied to data in particle physics by interpreting as images giving improved performances~\cite{ATLAS,Kasieczka_2017,Macaluso_2018,Lin:2018cin,andrews2019endtoend}. Image representations undergo difficulties in representing data from irregular geometry of detectors or due to sparsity of hit points. This results in the inherent loss of information concluding that image representations may constrain the amount of information which can be extracted.

Convolutional and Recurrent neural networks based approaches have deemed to be insufficient to address the challenges of realistic particle tracking~\cite{farrell2018novel}. Geometric Deep learning methods comprising of graph neural networks (GNN)~\cite{zhou2019graph} models, manifold learning models, etc which have demonstrated to be effective at characterizing tracks in realistic data~\cite{farrell2018novel,ju2020graph}. Graphs are constructed from the cloud of hits in each event, edges are drawn between hits that may come from the same particle track with some geometrical and physical constraints. GNN is then trained to classify the edges as real or fake, giving a true sample of track segments that can be used to construct a full track.

This works builds on the previous studies of GNNs applied to particle tracking, advancing in the regime of graph network construction and formulation, model performance, and full track reconstruction. This work is developed to be applied for tracking of cosmic ray muons originating from realistic extensive air showers in our designed 3-D scintillator detector modeled using GEANT4 simulation toolkit. This paper is organized as follows: Section~\ref{sec:algo} reviews the traditional algorithms currently used in track reconstruction,~\ref{sec:gdl} introduces the field of geometric deep learning and graph network formalism. Section~\ref{sec:detect} elucidates our modelled 3-D scintillator detector and its geometry. Section~\ref{sec:shower} describes how EAS simulations are performed over the detector. Section~\ref{sec:graph} describes our new GNN architecture developed for muon tracking. The results from our GNN model are shown in Section~\ref{sec:gnn}, with an extensive discussion over statistical analysis. The paper concludes with Section~\ref{sec:conc}.

\section{Traditional Track Reconstruction Algorithms}
\label{sec:algo}
Track reconstruction in HEP experiments is performed in several steps. In the first step, a set of hits from the detector is basically sampled at random as likely track candidates through hand-crafted criteria which are known as a seed. Secondly, track building is performed where additional hits are found by extrapolating the seeds using Kalman Filters~\cite{kalman} recursively while keeping the overall likelihood for a correct track in mind. Constrain over seed selection exist due to combinatorial explosion and it is necessary to select seeds that are likely to result in a successful track candidate.  One way is to select three hits (triplets) in the XY-plane perpendicular to the beam axis. An efficient initial seed generation can reduce time usage for such computationally intensive searches. The combinatorial nature of these algorithms has indicated that their computational cost will increase significantly with the high number of hit-points or increase in collision density.

\section{Geometric Deep Learning}
\label{sec:gdl}

Geometric Deep Learning comprises the application and modification of deep learning techniques to the graph, manifold, and unstructured data which can be represented in form of graphs. It can also be extended to apply to data having the underlying structure in a non-Euclidean space like social networks, regulatory networks in genetics, functional networks in brain imaging etc~\cite{9003285,gdl}.
The main focus is to learn functions on non-Euclidean structured domains to solve a variety of problems. Various algorithms have been posed to recover the lower dimensional structure i.e. manifold learning including different flavors of multidimensional scaling (MDS)~\cite{Tenenbaum2319}, locally linear embedding (LLE)~\cite{Roweis2323} etc. Recent approaches focus on embedding into graphs~\cite{Perozzi_2014,Tang_2015,mikolov2013efficient} and processing it by decomposing it into small sub-graphs.

Graph-structured data are omnipresent across various domains in science, engineering. The graph is defined as a set of nodes and edges of the network structure data. Graphs are the most natural and powerful way of representing unstructured data present in complex systems ~\cite{gdl,gilmer2017neural,zhou2019graph} e.g., the hierarchical structure of sentences and embeddings, protein structure network, social networks and relationships over time, etc. Graph Neural Networks (GNN's)~\cite{gdl, network} belongs to a niche of geometric deep learning architecture that implements learning functions that operate on graphs with relational inductive bias. A message-passing interface is performed whereby information is propagated across various nodes and edges in the graph allowing essential connections to originate and edge-, node-, and graph-level outputs to be computed. GNN's were firstly trained using Almeida-Pineda algorithm ~\cite{10,183787414c9f43cca64c0fc0ce6b3081} and applied for network analysis.

GNN has expanded its application to various domains in science and engineering. From molecular finger printing~\cite{Kearnes_2016} to Interaction networks~\cite{battaglia2016interaction} and graph networks to simulate increasingly complex physical systems~\cite{sanchezgonzalez2018graph,li2019learning,sanchezgonzalez2020learning} making a contribution and developing novel applications. Recent work~\cite{cranmer2019learning} has shown the applications of GNNs in extracting symbolic physical laws by learning message representations similar to the true force vector in n-body gravitational and spring-like simulations. GNs have also been applied to physics, adjusting their architectures by adding inductive biases to be consistent with Hamiltonian~\cite{sanchezgonzalez2019hamiltonian} and  Lagrangian mechanics~\cite{cranmer2020lagrangian}.

\subsection{Graph Network Formalism}
A brief review of the graph network (GN) formalism~\cite{network} which generalizes various scenarios of graph neural networks (GNNs), as well as other methods. A graph is generally represented by $G = (u,V,E)$, with $N_{v}$ as number of vertices (or nodes) and $N_{e}$ are the number of  edges. The V is the set of nodes (or vertices) i.e $V = \{v_{i}\}_{i=1:N_{v}}$ where $v_{i}$ are the attributes of $i$-th node. The \textbf{u} represents graph-level attributes. The E denotes set of edges or connection between vertices  $E = \{e_{n},r_{n},s_{n}\}_{i=1:N_{e}}$, where $e_{n}$ are the $n$-th edge attributes, $r_{n}$ and $s_{n}$ are receiver and sender indices of vertices to which edge is connected.

A GN block consists of update and aggregation functions as internal functions. The update functions are responsible for updating attributes of nodes or edges and they take fixed size input and output fixed-size output. The aggregation functions work on the link of various edges to one node by aggregating features from various nodes i.e. input a variable-sized set of inputs and output a fixed-size representation of input which is taken as input to the node.

A GN's processing stages, computation steps~\cite{zhou2019graph} and various functions are described as follows.
\begin{itemize}
    \item Edge Block: \begin{equation}
\label{eq:edge}
\begin{split}
e^{'}_{k} &= \phi^{e}(e_{k},v_{r_{k}},v_{s_{k}},u) \,
\qquad
\bar{e}^{'}_{i} = \rho^{e\xrightarrow[]{}v}(E^{'}_{i}) \,
\end{split}
\end{equation}
   \item Vertex Block: \begin{equation}
\label{eq:edge}
\begin{split}
v^{'}_{i} &= \phi^{v}(\bar{e_{i}},v_{i},u) \,
\qquad
\bar{e}^{'} = \rho^{e\xrightarrow[]{}u}(E^{'}) \,
\end{split}
\end{equation}
\item Global Block:\begin{equation}
\label{eq:edge}
\begin{split}
u^{'} &= \phi^{u}(\bar{e}^{'},\bar{v}^{'},u) \,
\qquad
\bar{v}^{'} = \rho^{v\xrightarrow[]{}u}(V^{'}) \,
\end{split}
\end{equation} 

\end{itemize}

where $\phi^{e},\phi^{v}, \phi^{u}$ are update functions and $\rho^{e\xrightarrow[]{}v},\rho^{e\xrightarrow[]{}u},\rho^{v\xrightarrow[]{}u}$ are aggregation functions. Usually ,the update functions are often implemented as Multi-layer perceptrons or deep neural networks. Aggregation functions are reduction operators like element-wise sums, means, or maximums which must be permutation invariant for permutation equivariance of GN block.  

The edge block computes the output attributes for each edge, $e^{'}_{k}$ using update functions, and aggregates them using the aggregation function at the receiving node, $e^{'}_{i}$, where $E^{'}_{i}$ corresponds to the edges incoming on the $i$-th node. The task of the vertex block is to compute one updated attribute for each node $v^{'}_{i}$. In the next step, the edge and node-level features are aggregated to update the global block or graph level attributes. The output of the formed Graph Network consist of all edge, node, and graph-level outputs, $G^{'}= (u^{'},V^{'},E^{'})$.

The above criteria define a generic GN block. These multiple GN blocks can be arranged in deep or recurrent configurations or by removing or rearranging internal components of the block such that information can be processed and propagated across the whole graph’s structure which enables to build of complex and long-range computations, relations between nodes, and vertices. This formalism is a general framework of a variety of models (like MPNN, NLNN ~\cite{gilmer2017neural,4700287}, and other variants) based on requirements and objectives that could be constructed by the GN framework. A popular GNN  architecture is Graph Convolutional Network (GCN)~\cite{kipf2017semisupervised} describing how to apply convolutions on the graph.

In the section~\ref{sec:graph}, we will describe our GN formalism based on the inspiration of generic GN block and implementation of the GNs internal functions w.r.t to the task we are aiming on. The choice of the specific architecture is motivated by element-wise relationships, the data type that exists in the input data (hit-points in our case), and the task we aim to solve.

\section{Detector Construction}
\label{sec:detect}
\begin{figure}[!hbt]
    \centering
    \includegraphics[scale=0.6]{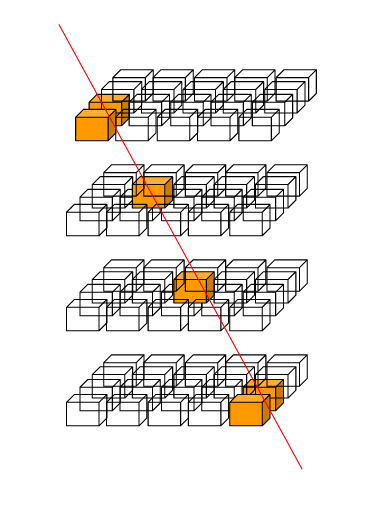}
    \caption{(color online) Passage of a particle (red) through grid of detector and depositing energy (yellow) in particular grid cells}
    \label{fig:part_det}
\end{figure}

Scintillator detectors are widely used to detect the passage of a particle through collecting and analyzing the light produced by the excitation of atoms and molecules. Their sensitivity to energy and fast response makes them the ideal detector to use. Usually, small cubical cells of scintillator are constructed and laid out in grids vertically as shown in Fig.~\ref{fig:part_det}. This provides us the spatial information of the particle with energy deposition also. However, when constructing a big detector it would require a marvelous amount of material, electronics, and hardware equipment which may increase the cost of production and development.
To optimize the number of electronics and other hardware used in the detector, we present a different approach to building a detector that will provide the same information about the passing particle. We constructed a 3-D Scintillator Detector of dimension $2m \times 2m \times 2 m$, consists of NaI scintillator tubes of dimension $0.9 cm \times 0.9 cm \times 2 m$ stacked horizontally and vertically alternative in each layer modeled using GEANT4 simulation toolkit~\cite{geant}. The combined horizontal and vertical layer corresponds to one Z layer. A gap of 1 mm is provided after each scintillator tube to resolve double hits in adjacent tubes. The horizontal layer will provide us the position on Y-axis where the vertical layer will provide us the position on the X-axis. 

\begin{figure}[!hbt]
    \centering
    \includegraphics[scale=0.55]{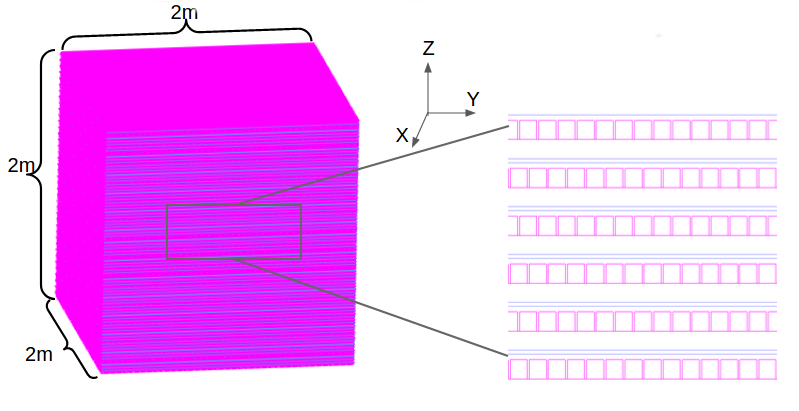}
    \caption{(color online) Schematic of detector Layout}
    \label{fig:total}
\end{figure}

An offset of 5 mm is provided in the alignment of tubes in each alternative Z layer to resolve the problem of particles passing through a 1 mm gap between tubes. An iron layer (blue color in the figure) of width 2 mm is provided after each layer and a 3 mm gap is given after each iron layer to curb the problem of saag present in the real-life detector. Visualization of the detector can be seen in Fig.~\ref{fig:total} There are a total of 80 Z layers combined or distinct 160 layers present in the detector having a total of 32000 NaI scintillator tubes. This technique resulted in the reduction of NaI tubes from  $200 \times 200 \times 200$ ( if cells of NaI were used) to 32000 tubes requiring less amount of electronics and hardware equipment.

\section{Air Shower Simulation}
\label{sec:shower}
Extensive Air Shower simulation is performed using CORSIKA (\textbf{CO}smic \textbf{R}ay \textbf{SI}mulations for \textbf{KA}scade)~\cite{corsika}. It is a Monte Carlo event generator for detailed simulation of extensive air showers initiated by high energy cosmic ray particles. The particles are tracked through the atmosphere until they undergo reactions with the air nuclei or - in the case of unstable secondaries - decay. The hadronic interactions between particles can be described by several reaction models. We have used the EPOS model which is based on the neXus framework. The location, magnetic field, and other location-specific parameters have been provided to simulate for our local place i.e IISER Mohali.

\begin{figure}[!hbt]
    \centering
    \includegraphics[scale=0.28]{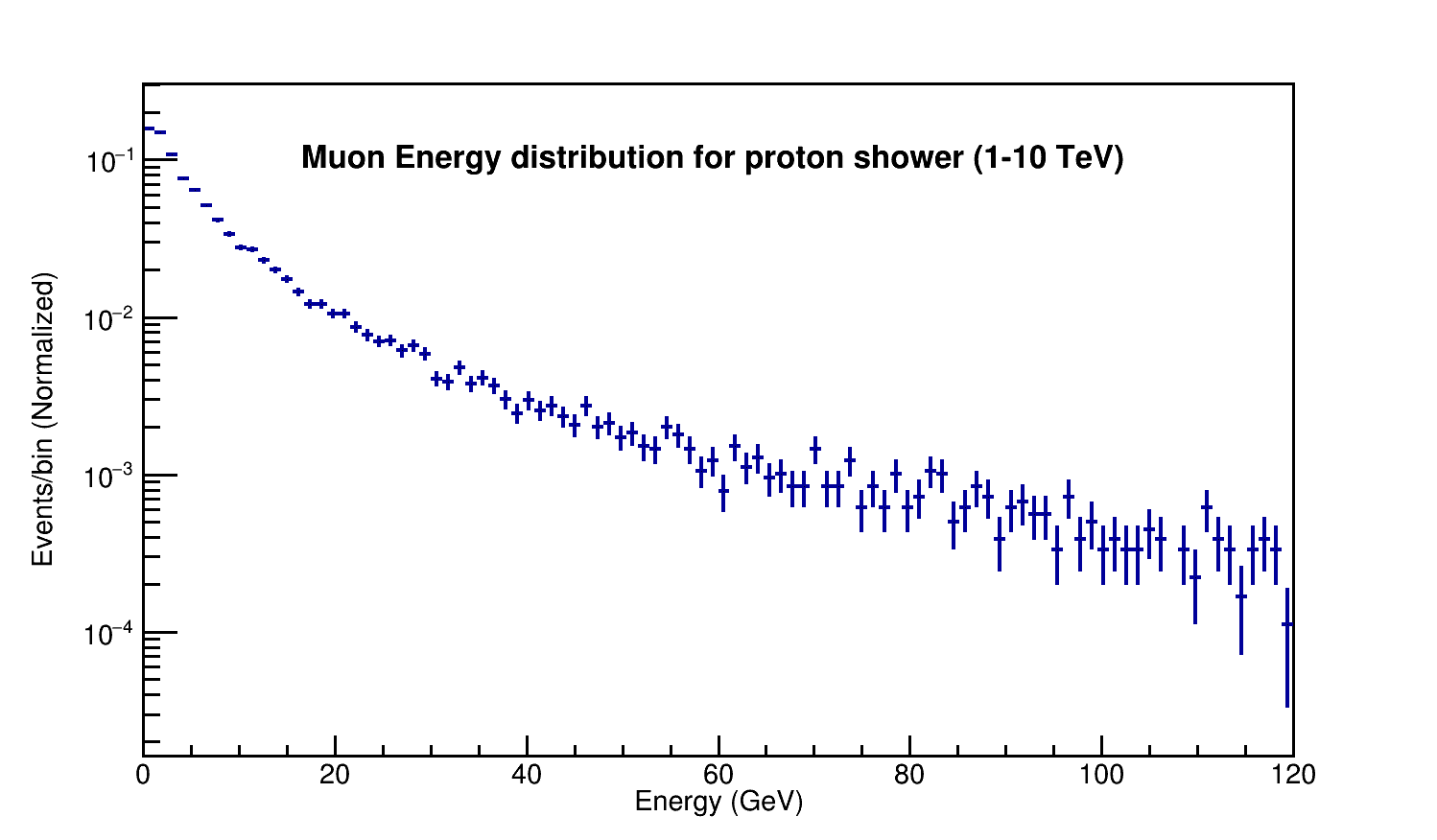}
    \caption{(color online) Muon Energy Spectrum from proton showers of energy range 1-10 TeV}
    \label{fig:mu_e_1_10}
\end{figure}

\begin{figure}[!hbt]
    \centering
    \includegraphics[scale=0.28]{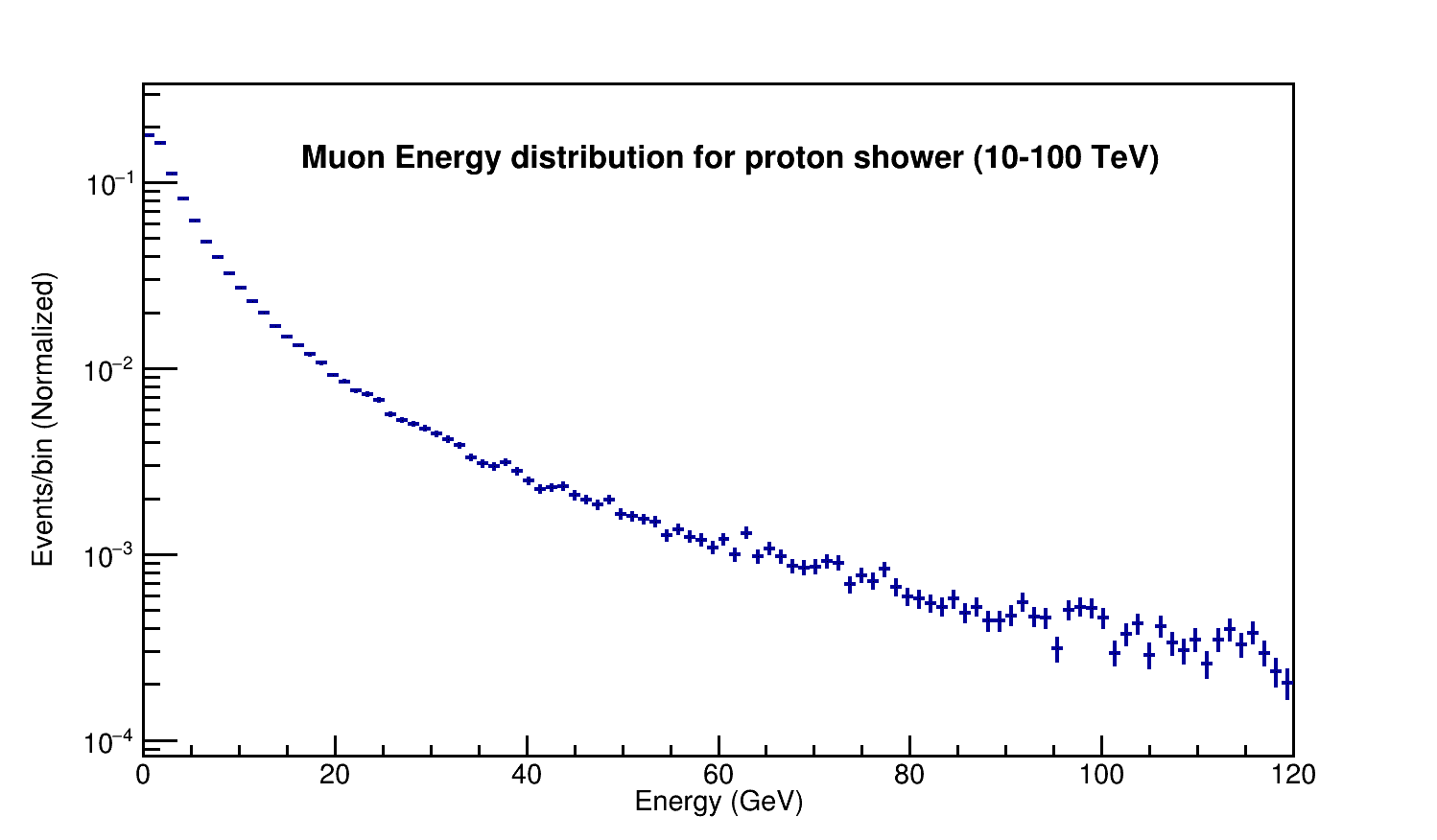}
    \caption{(color online) Muon Energy Spectrum from proton showers of energy range 10-100 TeV}
    \label{fig:mu_e_10_100}
\end{figure}

Proton showers are simulated having an initial particle energy range of 1-10 TeV and 10-100 TeV. Energy distribution of muons originating from these showers are shown in Fig.~\ref{fig:mu_e_1_10},~\ref{fig:mu_e_10_100}. We observe that low energy muons are abundant near the surface of the earth rather than high energy muons.

\begin{figure}[!hbt]
    \centering
    \hbox{\hspace{-3.0em}\includegraphics[scale=0.35]{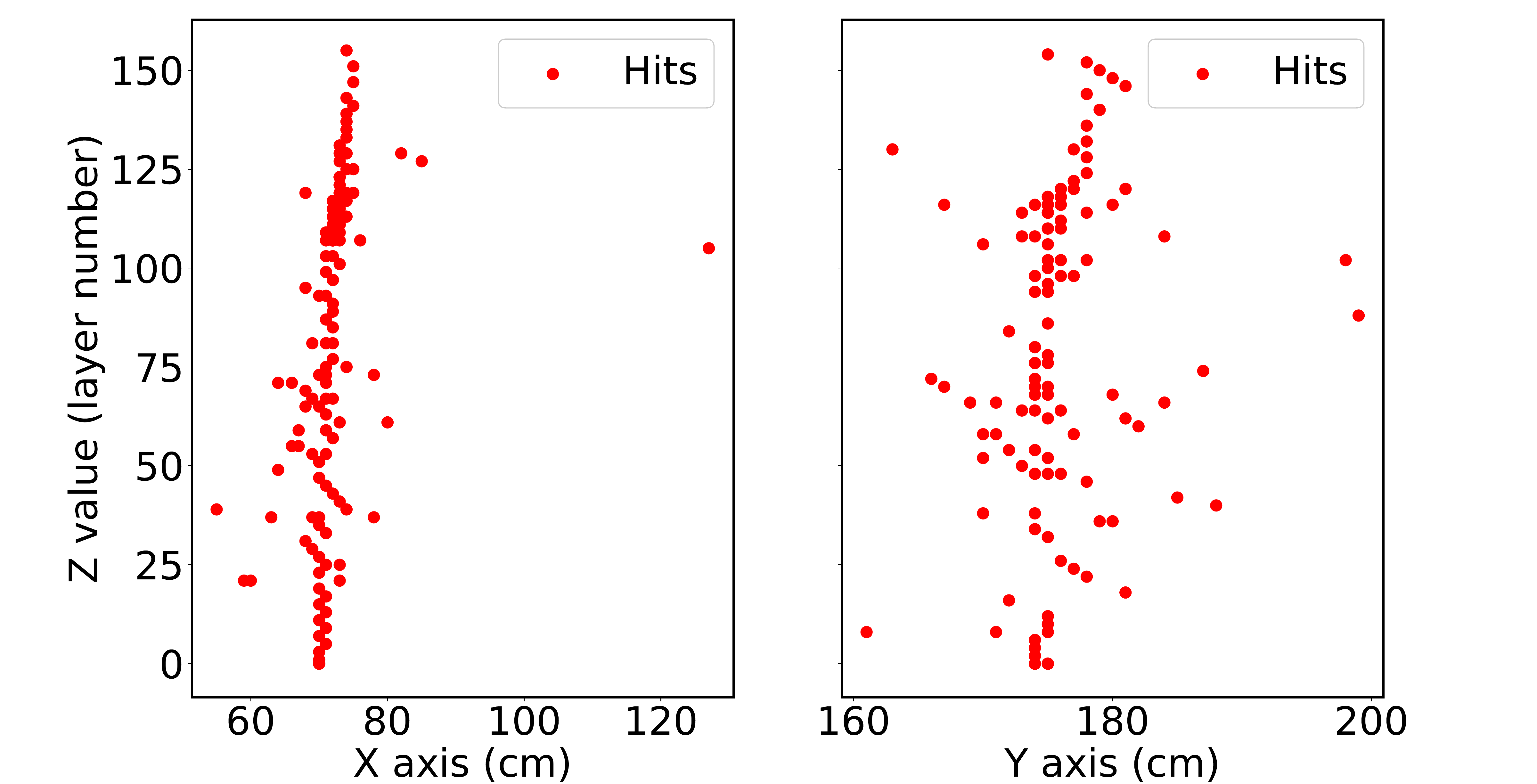}}
    \caption{(color online) Hit points of muon passing through scintillator detector}
    \label{fig:muon}
\end{figure}

\begin{figure}[!hbt]
    \centering
    \hbox{\hspace{-2.0em}\includegraphics[scale=0.8]{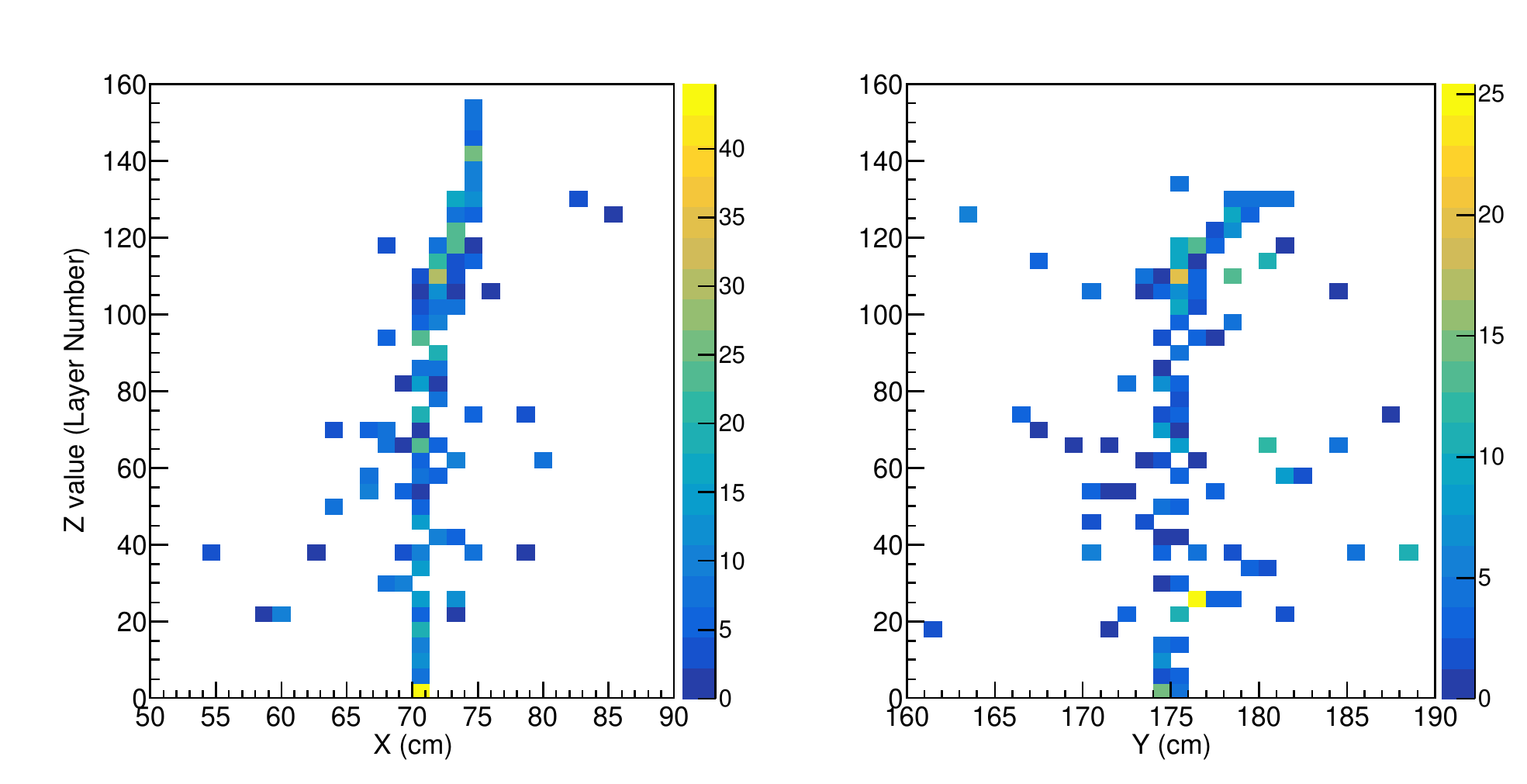}}
    \caption{(color online) Event Visualization as Energy deposited(MeV) represented by colorbar by muon in the detector.}
    \label{fig:interaction}
\end{figure}

CORSIKA is interfaced with GEANT4 by projecting muons having energy sampled from these distributions from top of the detector as shown in Fig.~\ref{fig:total} in a random x,y position. Muons then interact with the detector depositing energy in the tubes and providing particle hit-points in the detector. These hit-points provide us the track information of the primary particle as well as the secondary particles generated during the interaction. Hitpoints measurements in X and Y direction along with Z value (layer number) for a muon particle passing through scintillator detector is shown in Fig.~\ref{fig:muon}. The interaction view of muon (event visualization) with the detector is represented in Fig.~\ref{fig:interaction} as a function of energy deposited by muon in the scintillator.

\section{Graph Construction}
\label{sec:graph}
\begin{figure}[!hbt]
    \centering
    \hbox{\hspace{-0.1em}\includegraphics[scale=0.4]{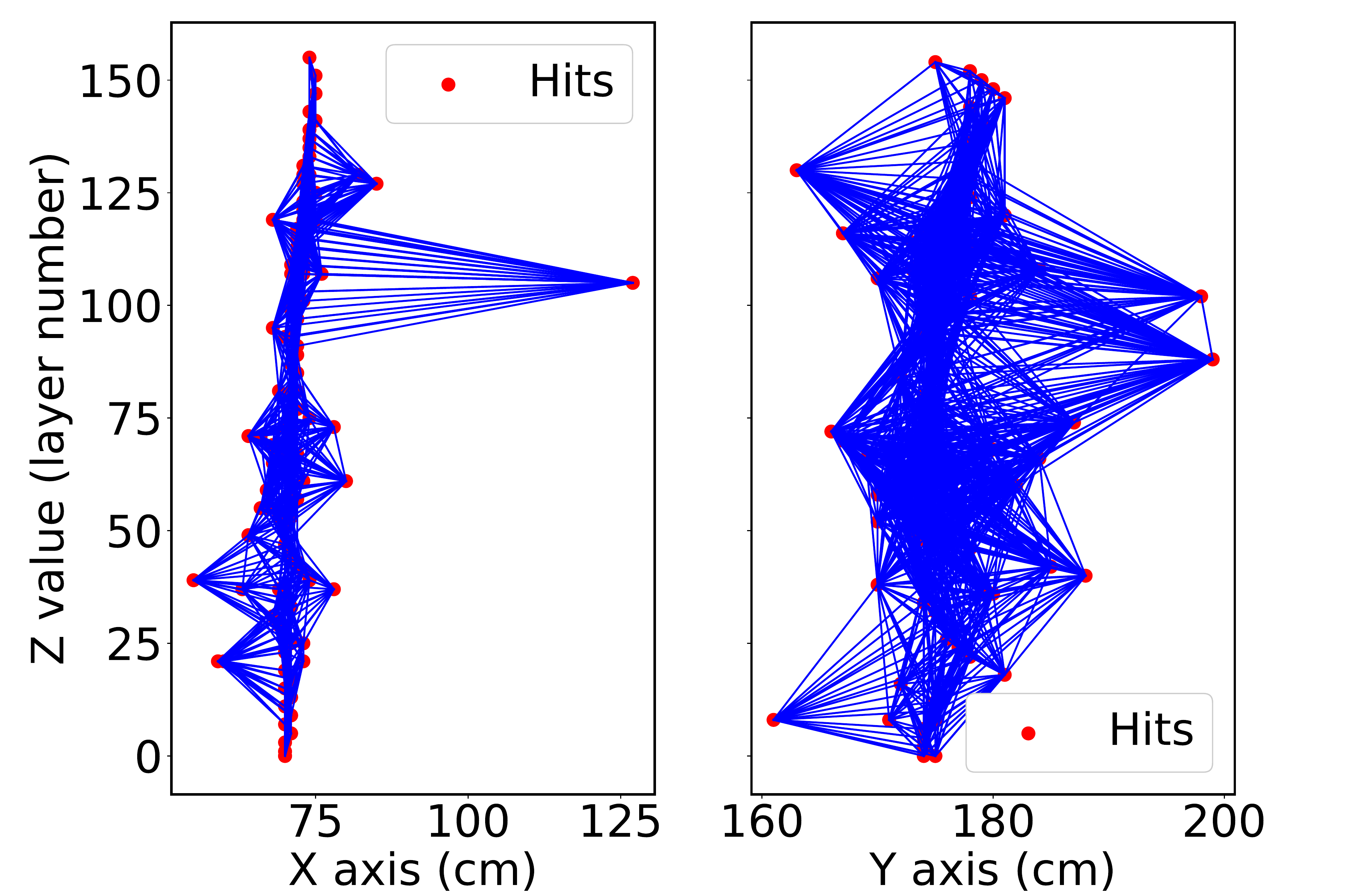}}
    \caption{(color online) Graph constructed from hit points in our 3-D scintillator detector. Hit points (red) are nodes of graph and blue lines are edges between nodes. }
    \label{fig:graph}
\end{figure}

The nature of the relationships between different elements in the input data optimizes the construction of graph and pairwise networks between nodes. Different graph construction methods can be implemented depending on the task~\cite{graph_arch} resulting in complex graph architecture to simple architecture. An example of graph construction of nodes and edges is shown in Fig.~\ref{fig:graph}\\
Edges in the graph form the basis of communication and connection channels among the nodes. Moreover, these features can indicate an association between different data inputs and can encode physics motivating variables. For small size input data, it is beneficial to form a fully connected graph, which enables the network to learn which object connections are important.  In larger sets, fully connected graph will lead to increase in the number of edges between all nodes, the computational load of creating numerous neural network for edge representation becomes prohibitive. Dynamic graph construction, where node features can also be  based on a learned representation and is implemented in EdgeNet~\cite{edge}, GraveNet~\cite{grave} architectures.\\
We present a general graph construction approach where the edge formation between hit-points is restricted by various geometric and physical constraints. In doing so, we first encode our hit measurements into an embedded space using a  multi-layer perceptron (MLP) to find the best representation for them. We have developed a tracking application using Graph Neural Networks (GNN) to identify the track of primary particle in a partially-labeled graph by binary classifying the graph nodes. The inputs to the model are the hit coordinates (node features) and the connectivity specification.

\subsection{GNN Architecture}
We developed a tracking algorithm based on the MPNN network formerly described in~\cite{kipf2017semisupervised,choma2020track} with a similar attention mechanism described in~\cite{2018graph} and an operation between message passing layers to help reduce the vanishing gradient effect. We have modified the previous approaches towards the MPNN network to apply them to our problem of muon tracking in a 3-D scintillator detector.  Once hits are organised into input graphs $G_{in}$, the hit coordinates of $i$-th vertex (($x_{i},z_{i}$) vertically stacked with ($y_{i},z_{i}$)) are encoded into a latent space using a multi-layer perceptron (MLP) with trainable parameters $\theta$ (Encoder Network). The MLP consists of 3 layers of a fully connected neural network connecting the input hit-points to a latent representation of them. \\
We then introduce a recurrent set of $\{$Connection Network, Vertex Network$\}$ iterating $n_{iter}$ times over the connection and vertex network sequentially. The latent inputs after the Encoder Network are passed through Connection Network (MLP) after concatenation with the features of the connecting vertex. The output of the Connection network is then used to calculate the score for each connection through the concatenation of latent features of various connections. This score is further used in an attention mechanism as a message passing between subsequent vertex with the vertex features. The MLP for Connection Network and Vertex Network consists of 3 layers having 32 nodes with Layer Normalization each with ReLU activation function. \\
The function of Vertex Network is to execute a message-passing forward pass, for each vertex, the neighboring vertex features of all incoming and outgoing connections are summed i.e. each vertex receives the sum of the messages of all connecting connections. Then this is passed through a vertex network MLP, after the addition of hidden features to the previous iteration is executed to preserve information between messages. After $n_{iter}$ iterations, the collective connection features are passed through the Connection Classification network of Multi-layer perceptron constructed of fully connected dense layer network a final time, to determine the final connection scores. They are interpreted as a truth likelihood describing the probability of a connection (track in or case) to be fake or true and handled by a binary cross-entropy loss function.
\begin{figure}[!hbt]
    \centering
    \includegraphics[width=\textwidth]{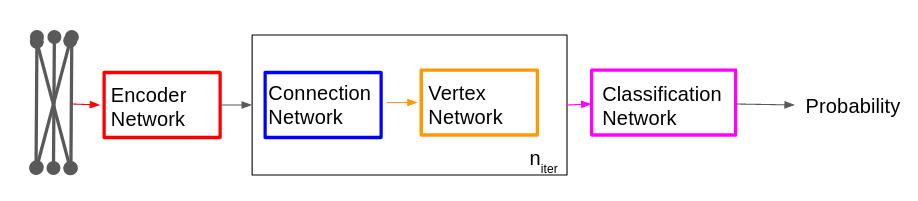}
    \caption{(color online) Flow of Information in our Graph Network}
    \label{fig:my_label}
\end{figure}
\subsection{Graph Hit Classification}
\label{sec:gnn}
The graph hit classification model performs binary classification of the connections of the graph between vertexes to specify the primary particle track vs fake track. The graphs are constructed by considering hit points as the vertex of the graph and constructing connections between the vertex underneath geometrical and physical constraints connecting all hits on adjacent layers. The model uses four graph recursions of network which is followed by a final classification layer that predicts whether the connection belongs to the primary particle track or not.

Given the above architecture, we present the results of connection classification to find tracks for the primary particle i.e. muon. We present a hierarchical application of our graph network starting from single track finding to the multi-track finding of multiple cosmic ray muons passing through the detector. 

\subsubsection{Single Track Finding}
\begin{figure}[!hbt]
\def\tabularxcolumn#1{m{#1}}
\begin{tabularx}{\linewidth}{@{}cXX@{}}
\begin{tabular}{cc}
\subfloat[Training Loss]{\hspace{-1.0em}\includegraphics[scale=0.41]{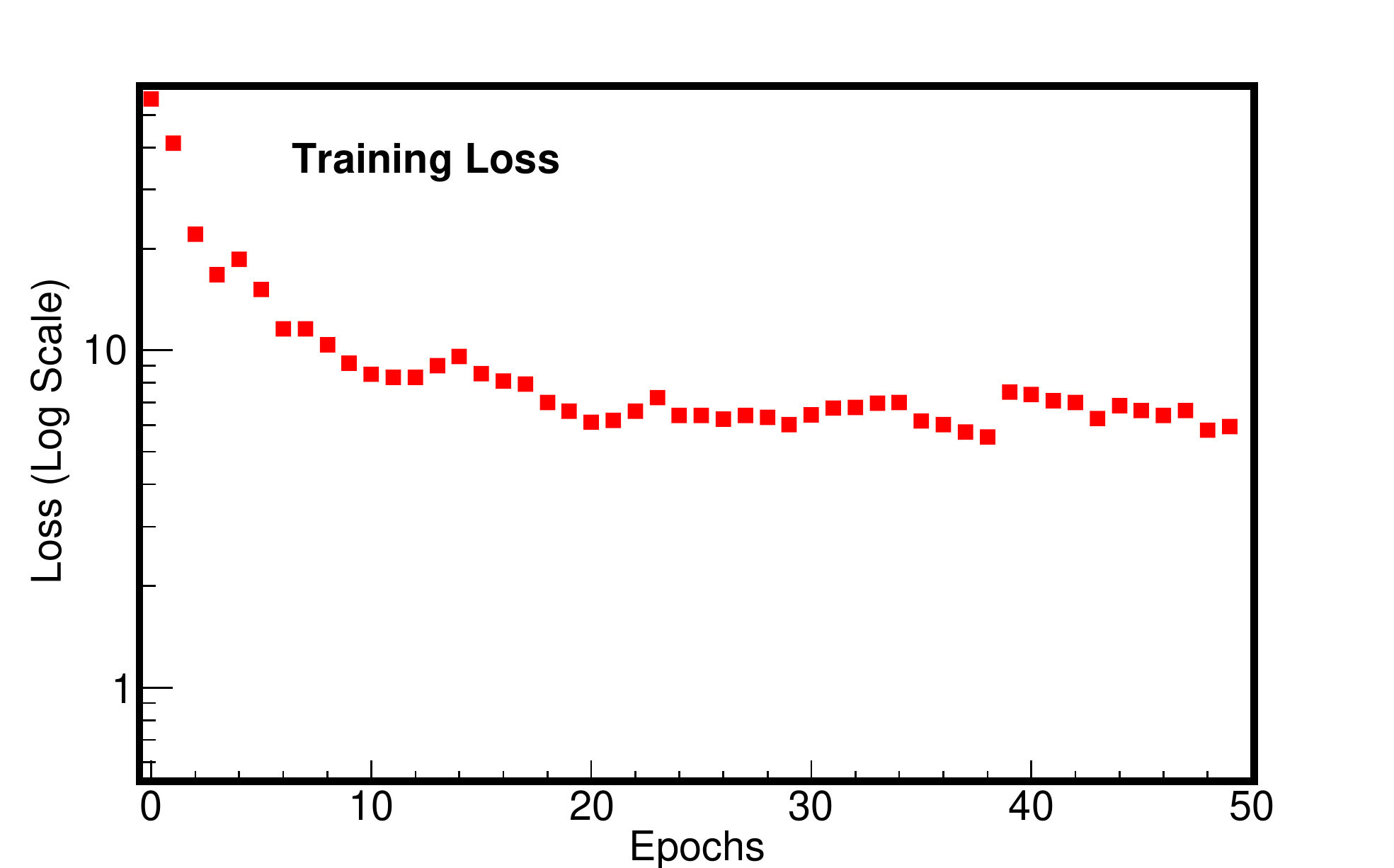}} 
   & \subfloat[Validation Loss]{\hspace{-2.0em}\includegraphics[scale=0.41]{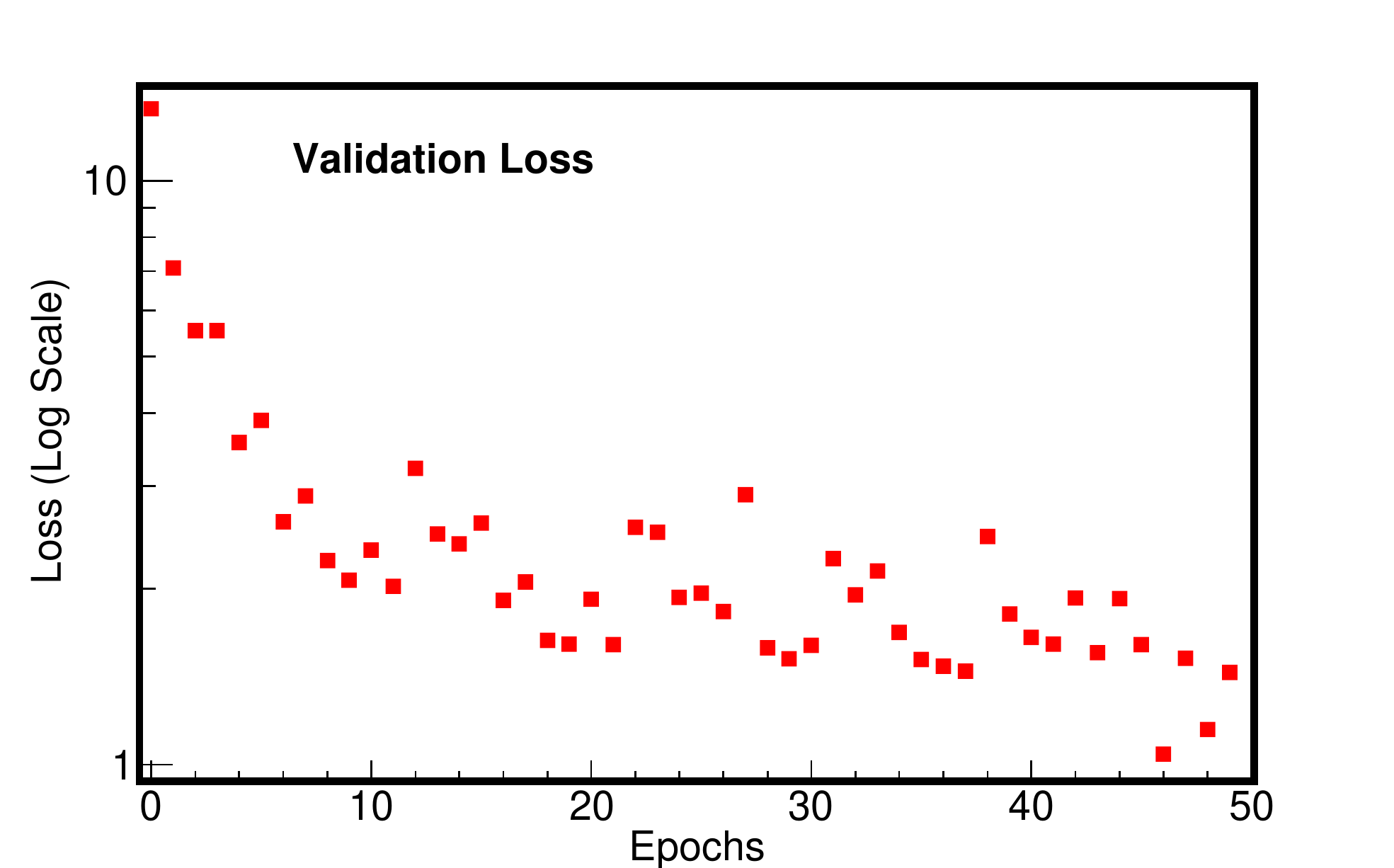}}\\
\end{tabular}
\end{tabularx}
\caption{(color online) Binary Cross-Entropy Loss vs Epochs for Graph Neural Network}\label{fig:loss}
\end{figure}

\begin{figure}[!hbt]
\def\tabularxcolumn#1{m{#1}}
\begin{tabularx}{\linewidth}{@{}cXX@{}}
\begin{tabular}{cc}
\subfloat[Training Accuracy]{\hspace{-1.0em}\includegraphics[scale=0.41]{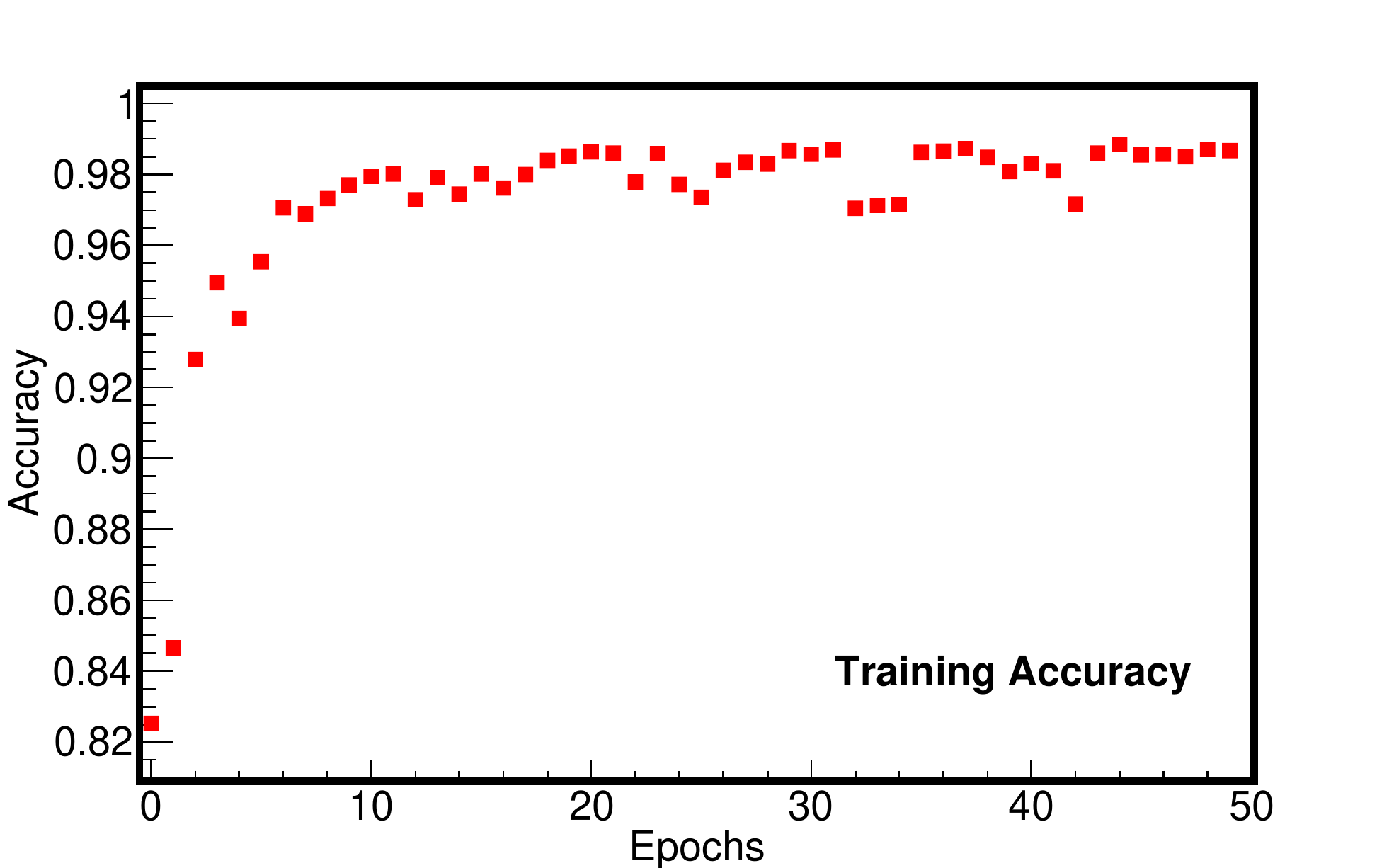}} 
   & \subfloat[Validation Accuracy]{\hspace{-2.0em}\includegraphics[scale=0.41]{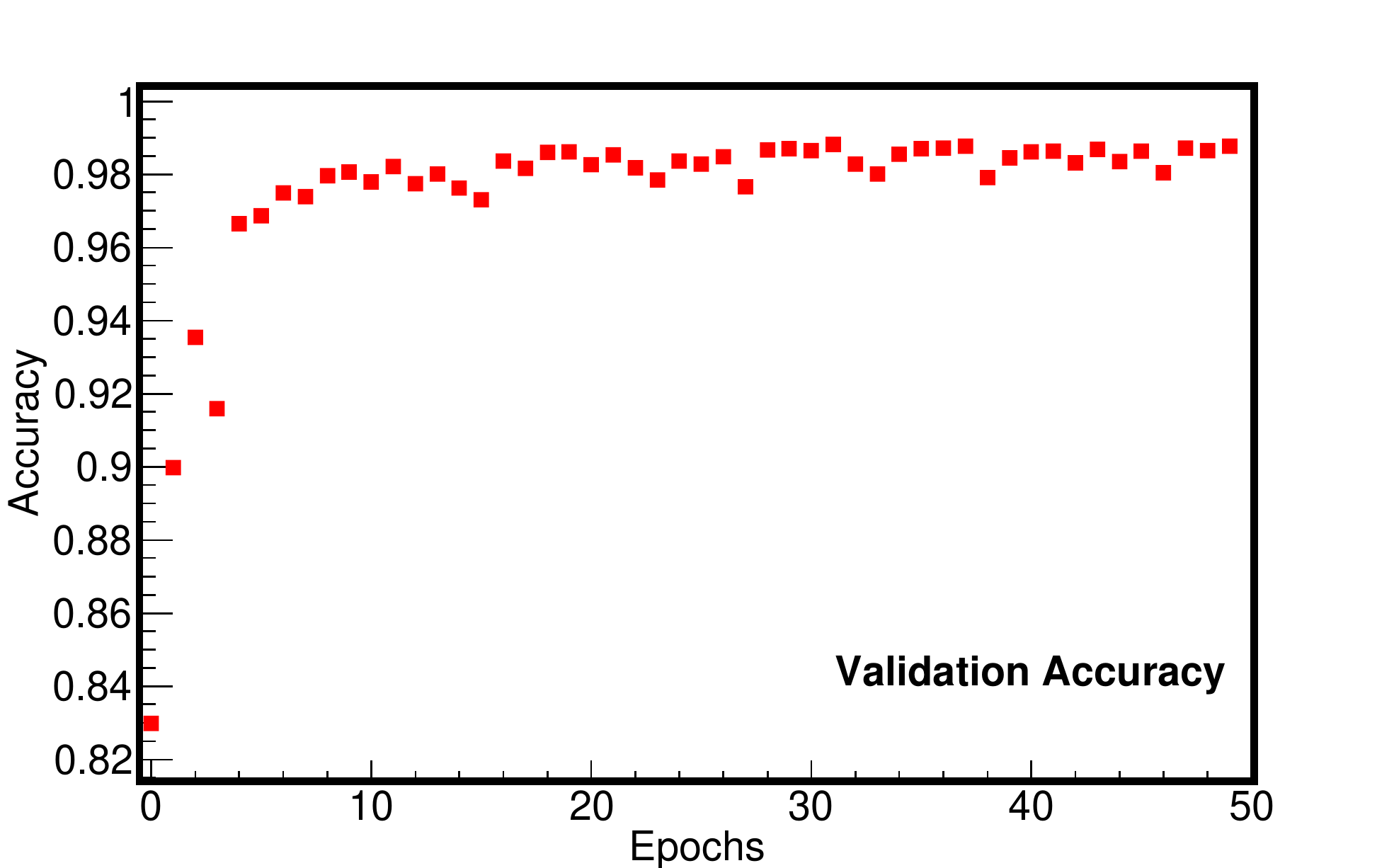}}\\
\end{tabular}
\end{tabularx}
\caption{(color online) Accuracy vs Epochs for Graph Neural Network}\label{fig:acc}
\end{figure}

Single track finding aims to reconstruct a single primary particle track from the hit-points in the detector (Fig.~\ref{fig:muon}). Data was collected by simulating the passage of a single muon particle in the detector. 
Results for the model loss vs epochs trained for single track finding is shown in Fig.~\ref{fig:loss} and model accuracy vs epochs is shown in Fig. ~\ref{fig:acc}.

We infer from Fig.~\ref{fig:loss} and Fig.~\ref{fig:acc} that model is able to encode the representation of the detector and able to find the true tracks corresponding to the primary particle. A validation and training accuracy of $\sim$ 98\% is obtained when the model is trained for 50 epochs in a batch size of 100. The model seems to converge after 50 epochs as training loss and accuracy do not vary significantly after 40 epochs. The difference between the scale of validation loss and training loss is due to the number of samples used for training and validation of the model. 
%\begin{figure}[!hbt]

\subsubsection{Dual Track Finding}
\begin{figure}[!hbt]
\def\tabularxcolumn#1{m{#1}}
\begin{tabularx}{\linewidth}{@{}cXX@{}}
\begin{tabular}{cc}
\subfloat[Training Loss]{\hspace{-1.0em}\includegraphics[scale=0.41]{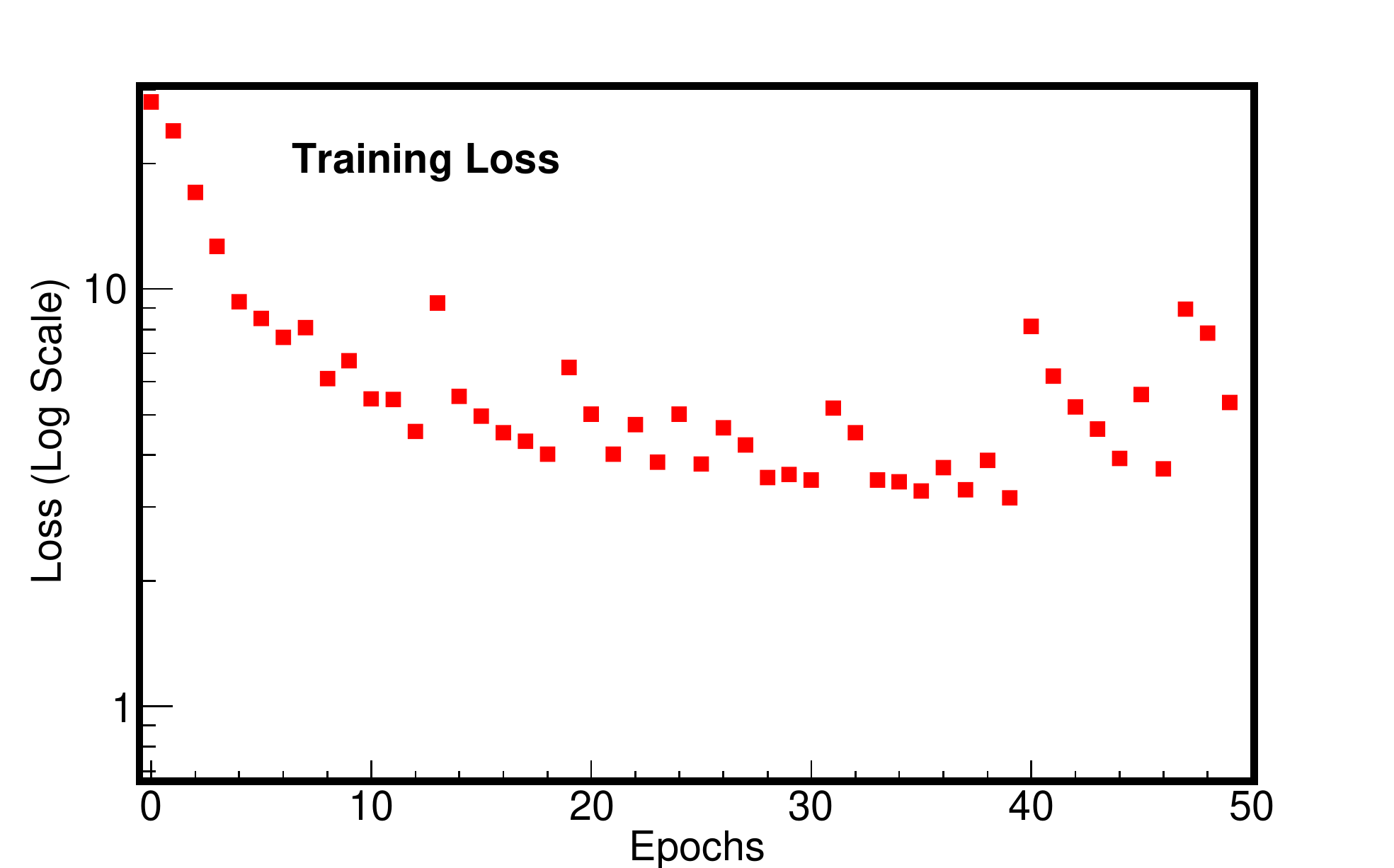}} 
   & \subfloat[Validation Loss]{\hspace{-2.0em}\includegraphics[scale=0.41]{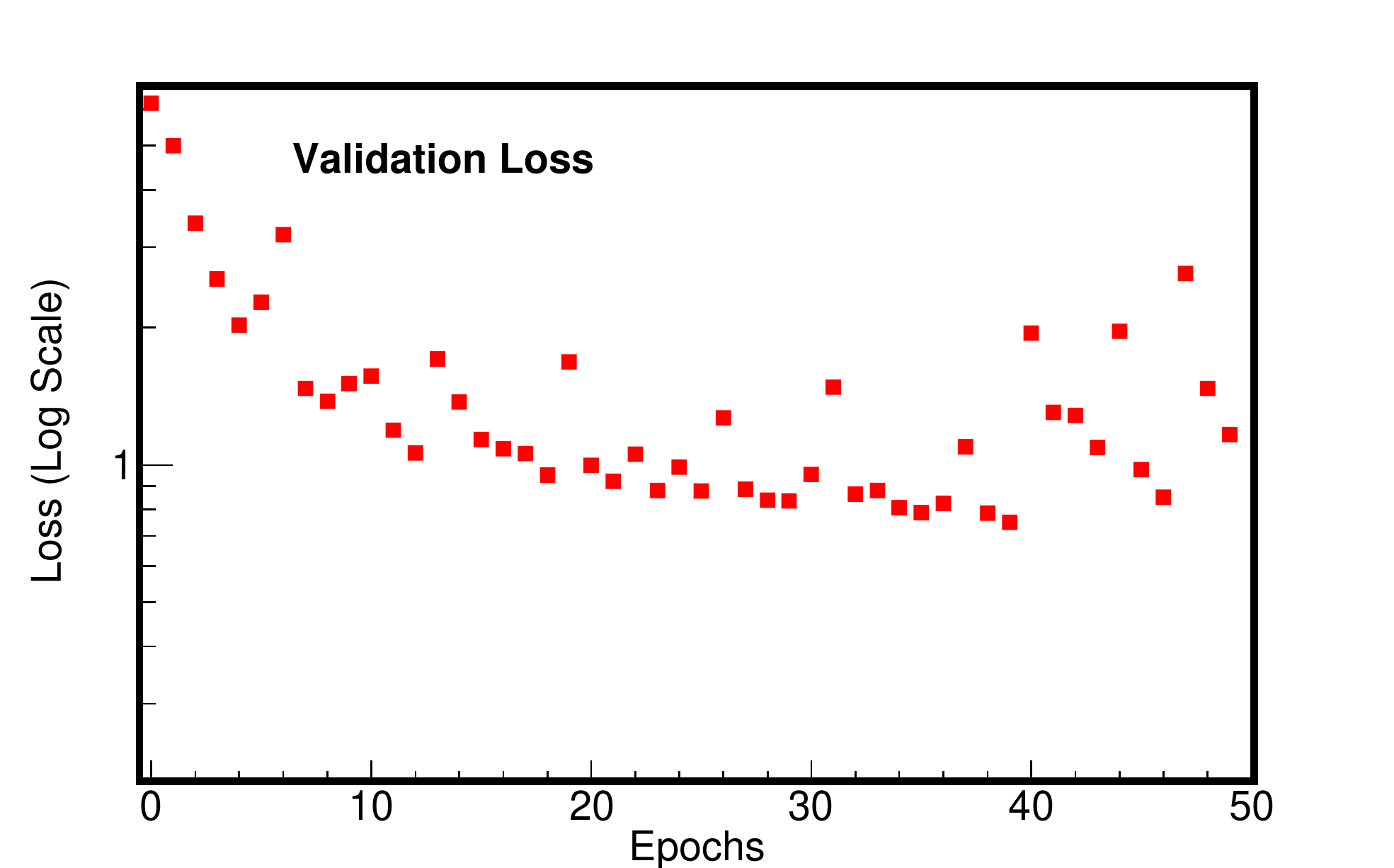}}\\
\end{tabular}
\end{tabularx}
\caption{(color online) Binary Cross-Entropy Loss vs Epochs for Graph Neural Network}\label{fig:loss1}
\end{figure}

\begin{figure}[!hbt]
\def\tabularxcolumn#1{m{#1}}
\begin{tabularx}{\linewidth}{@{}cXX@{}}
\begin{tabular}{cc}
\subfloat[Training Accuracy]{\hspace{-1.0em}\includegraphics[scale=0.41]{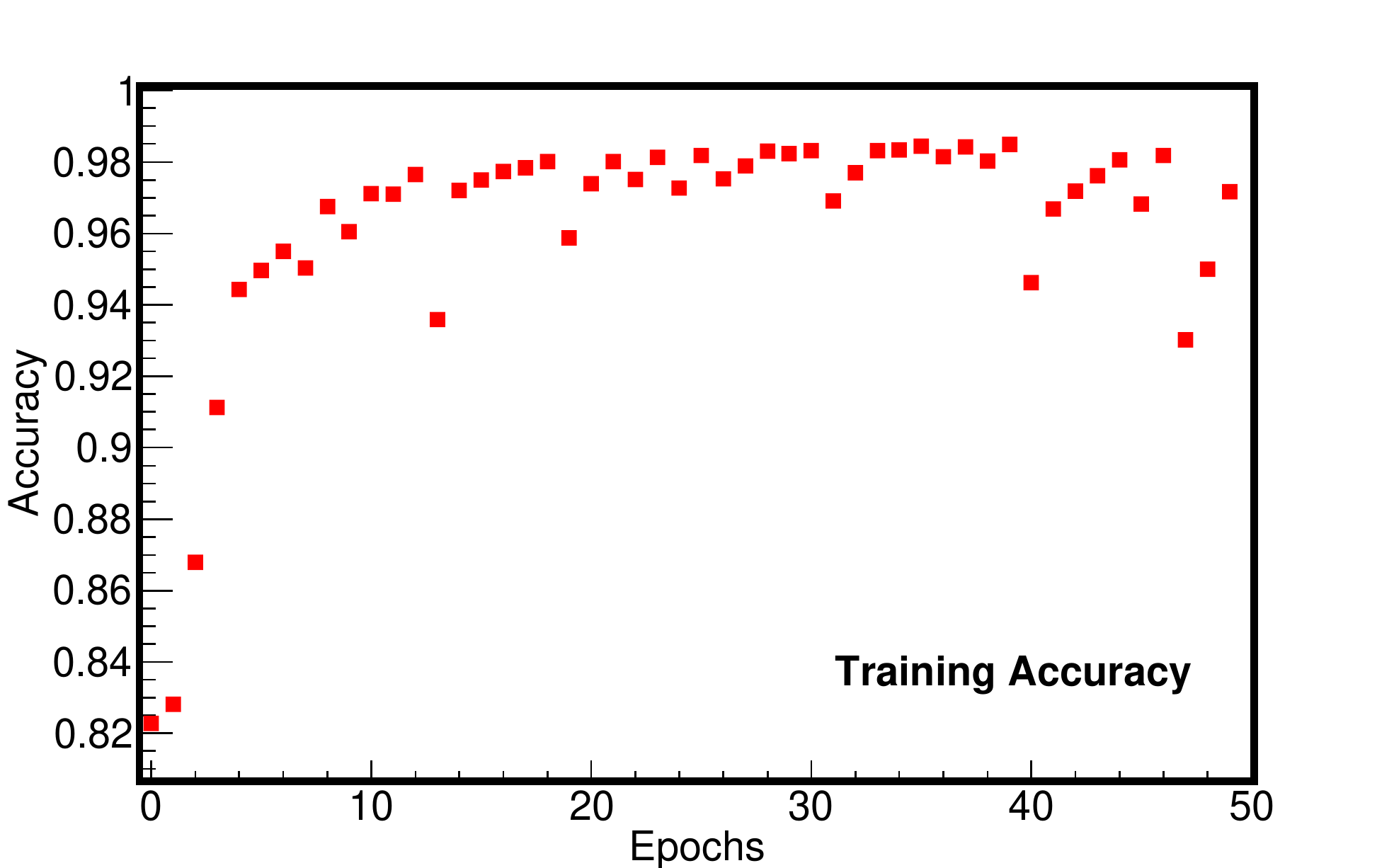}} 
   & \subfloat[Validation Accuracy]{\hspace{-2.0em}\includegraphics[scale=0.41]{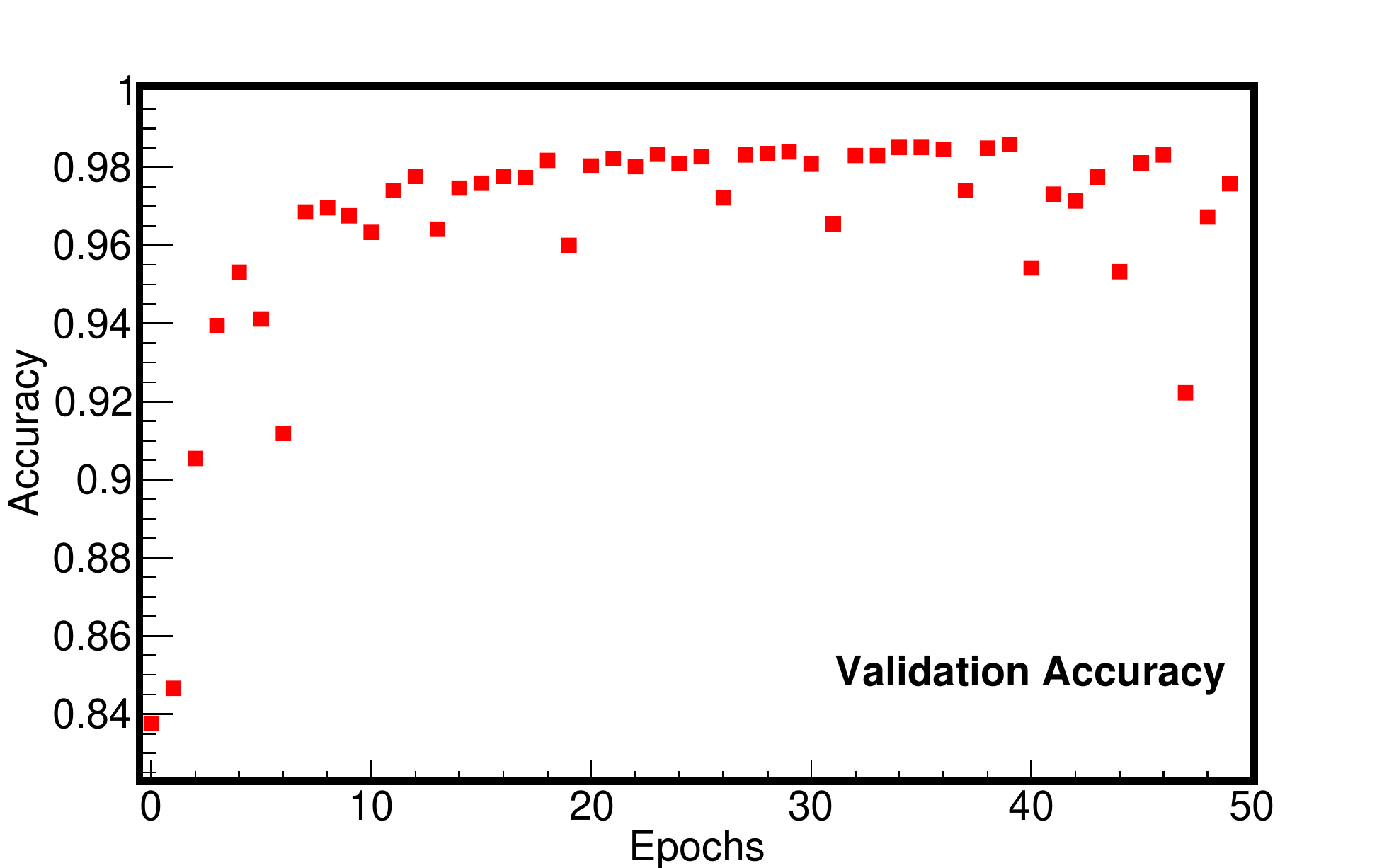}}\\
\end{tabular}
\end{tabularx}
\caption{(color online) Accuracy vs Epochs for Graph Neural Network}\label{fig:acc1}
\end{figure}

We extend the application of our Graph Network to find dual particle tracks of primary particle passing through detector eg. two muons passing through detector simultaneously. Data was generated by passing 2 muons in random directions through the detector. The results for the model loss vs epochs is shown in Fig.~\ref{fig:loss1} and model accuracy vs epochs is shown in Fig.~\ref{fig:acc1} 

The model is able to perform well in finding tracks of two muons passing through the detector with an accuracy of $\sim$ 97\%. The training loss converges smoothly to the minimum with a few jumps in between which may occur due to some events which are hard to learn. Similar behavior in accuracy w.r.t single track finding is observed here. Moreover, the graph network model is robust towards the prevailing noise like double hits, duplicate hits in the detector to find the primary track.

\subsubsection{Multi Track Finding}
\begin{figure}[!hbt]
\def\tabularxcolumn#1{m{#1}}
\begin{tabularx}{\linewidth}{@{}cXX@{}}
\begin{tabular}{cc}
\subfloat[Training Loss]{\hspace{-1.0em}\includegraphics[scale=0.41]{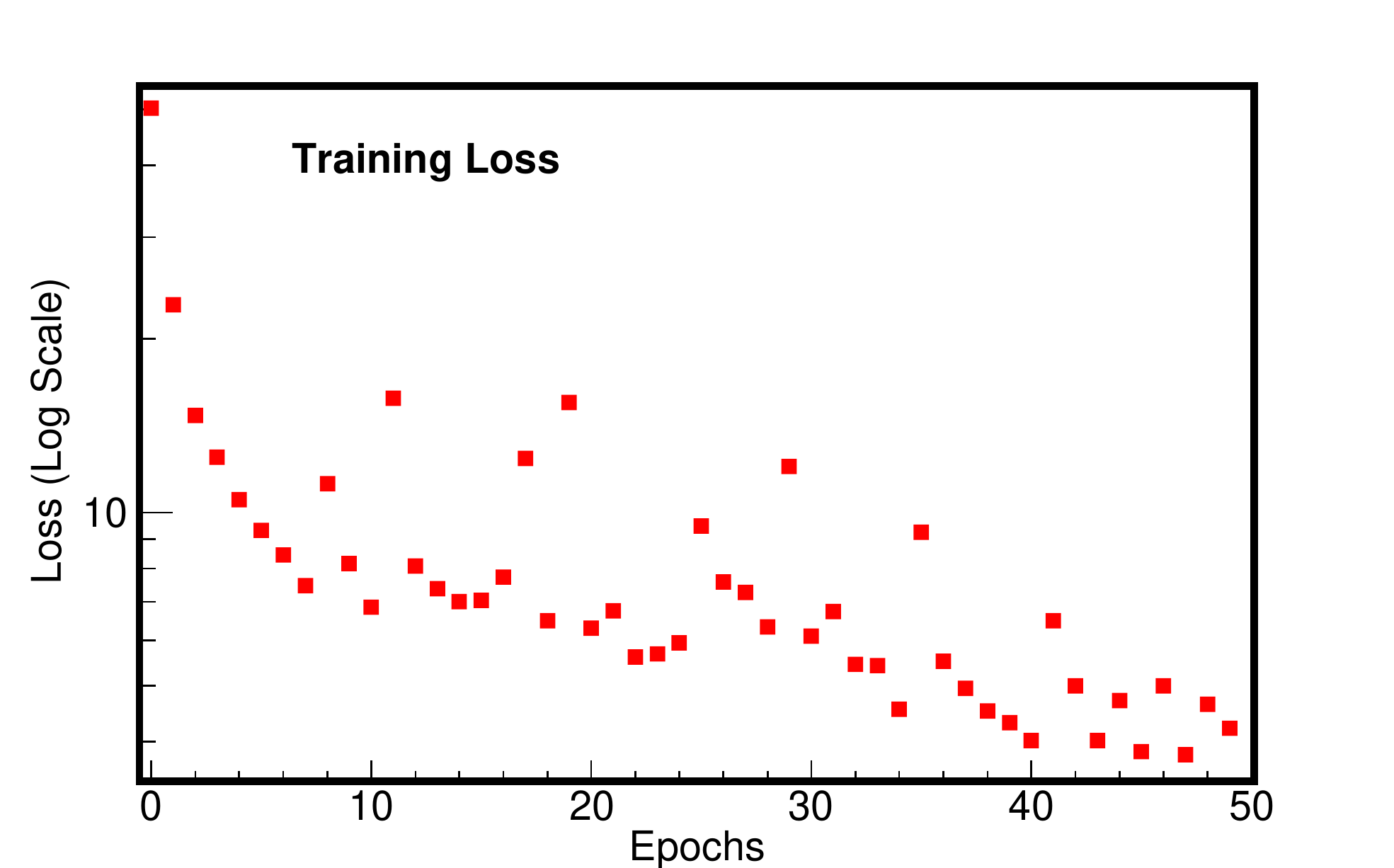}} 
   & \subfloat[Validation Loss]{\hspace{-2.0em}\includegraphics[scale=0.41]{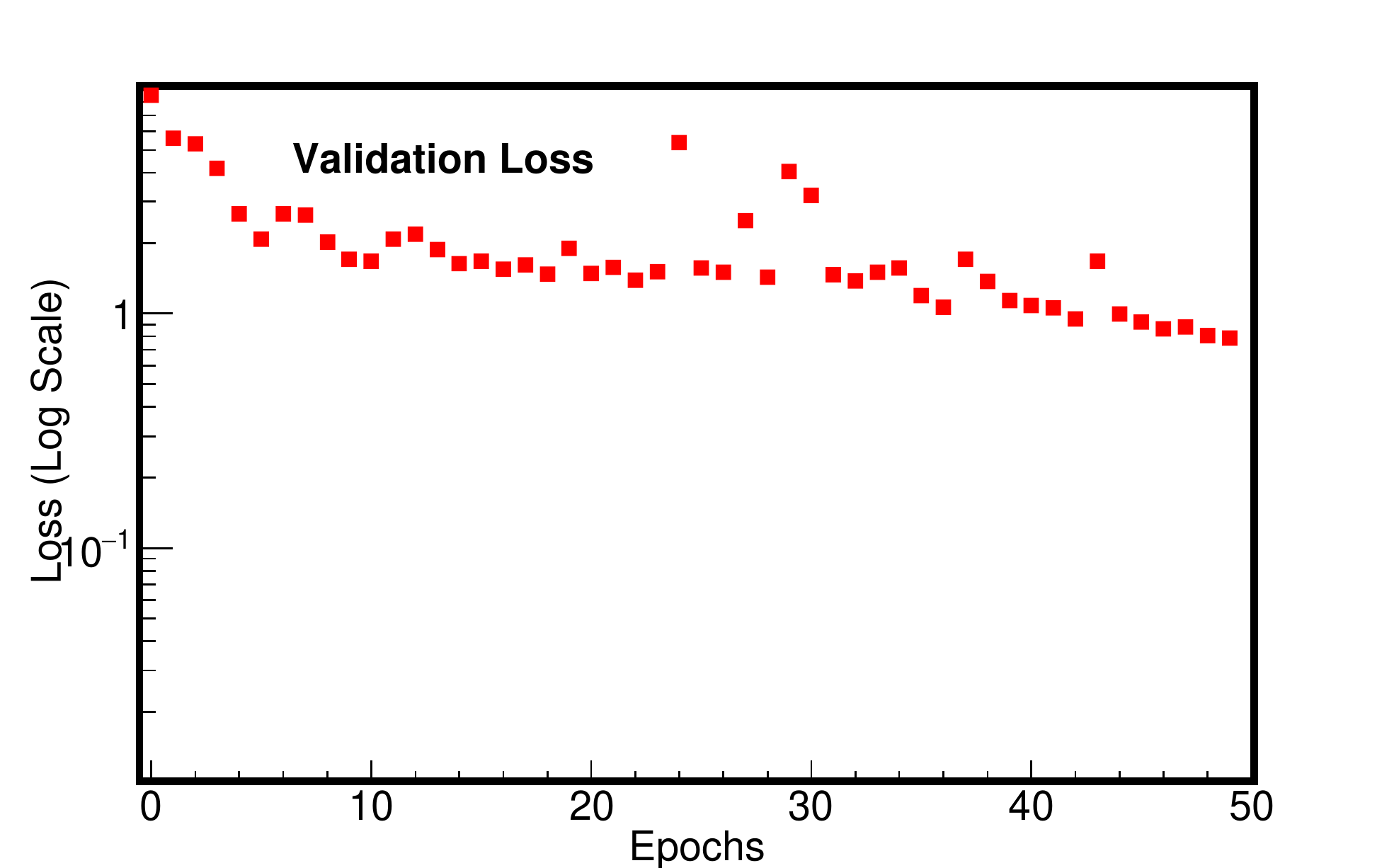}}\\
\end{tabular}
\end{tabularx}
\caption{(color online) Binary Cross-Entropy Loss vs Epochs for Graph Neural Network}
\label{fig:loss2}
\end{figure}

\begin{figure}[!hbt]
\def\tabularxcolumn#1{m{#1}}
\begin{tabularx}{\linewidth}{@{}cXX@{}}
\begin{tabular}{cc}
\subfloat[Training Accuracy]{\hspace{-1.0em}\includegraphics[scale=0.41]{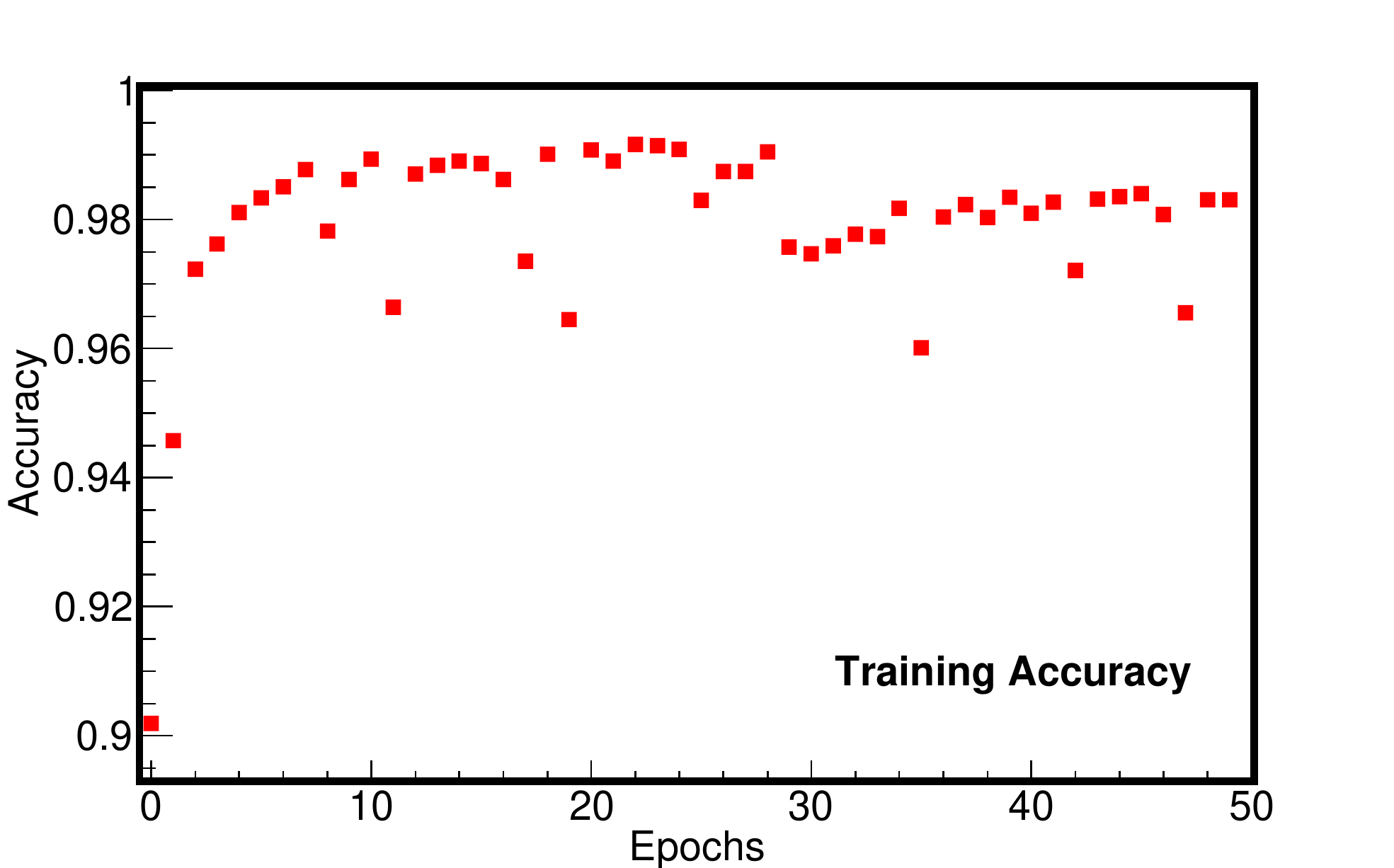}} 
   & \subfloat[Validation Accuracy]{\hspace{-2.0em}\includegraphics[scale=0.41]{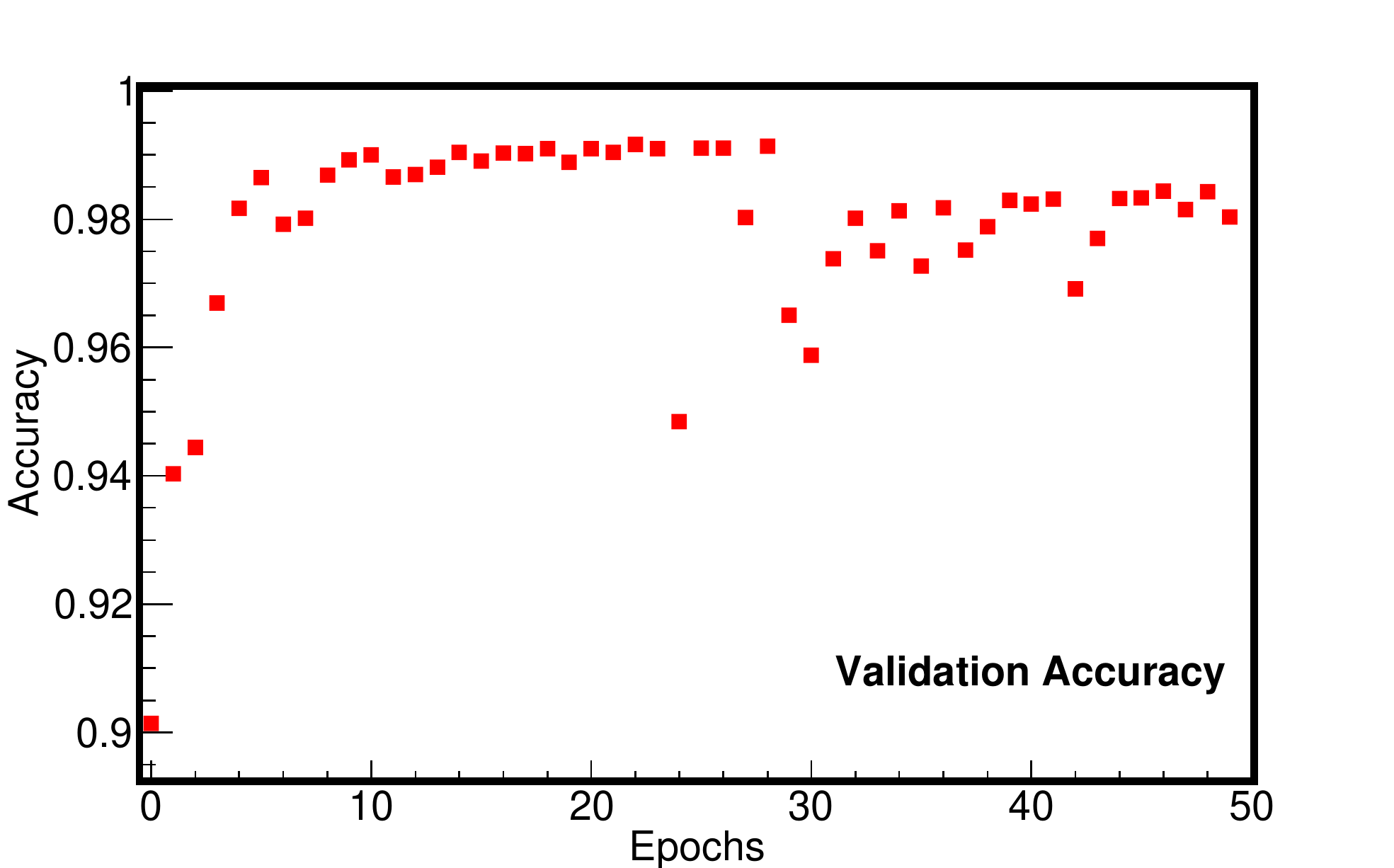}}\\
\end{tabular}
\end{tabularx}
\caption{(color online) Accuracy vs Epochs for Graph Neural Network}\label{fig:acc2}
\end{figure}

After, getting a good performance over single and dual-track finding. We further extend our range to multi-track finding where the number of tracks is restricted to the conditions (Extensive Air Showers) to which we are applying our algorithm. Each event will consist of a random number of tracks where the minimum is one track and maximum tracks are restricted by the conditions. Results for the model loss vs epochs is shown in Fig.~\ref{fig:loss2} and model accuracy vs epochs is shown in Fig.~\ref{fig:acc2}.

The maximum number of tracks possible in an event for this model was set to eight tracks. The Fig.~\ref{fig:acc2} and Fig.~\ref{fig:loss2} provides a positive feedback about the performance of graph network for its application in multi-track finding. We achieved an accuracy of about $\sim$ 98\% to find the particle tracks.

\subsection{Model Diagnostics}
Accuracy is not the sole measure of the performance of the model. To further justify the performance of our model, we performed various diagnostics like ROC Curve, etc.  The Receiver Operating Characteristic (ROC) curve is a graph showing the performance of a classification model which gives an area under curve (AUC) parameter. The Area Under the Curve is the measure of the ability of a classifier to distinguish between classes. It shows that model can distinguish between fake and true hits very well. This diagnostic test is performed for all the models in the single track, dual track, and multi-track finding scenarios. 

\begin{figure}
    \centering
    \includegraphics[width=0.32\textwidth]{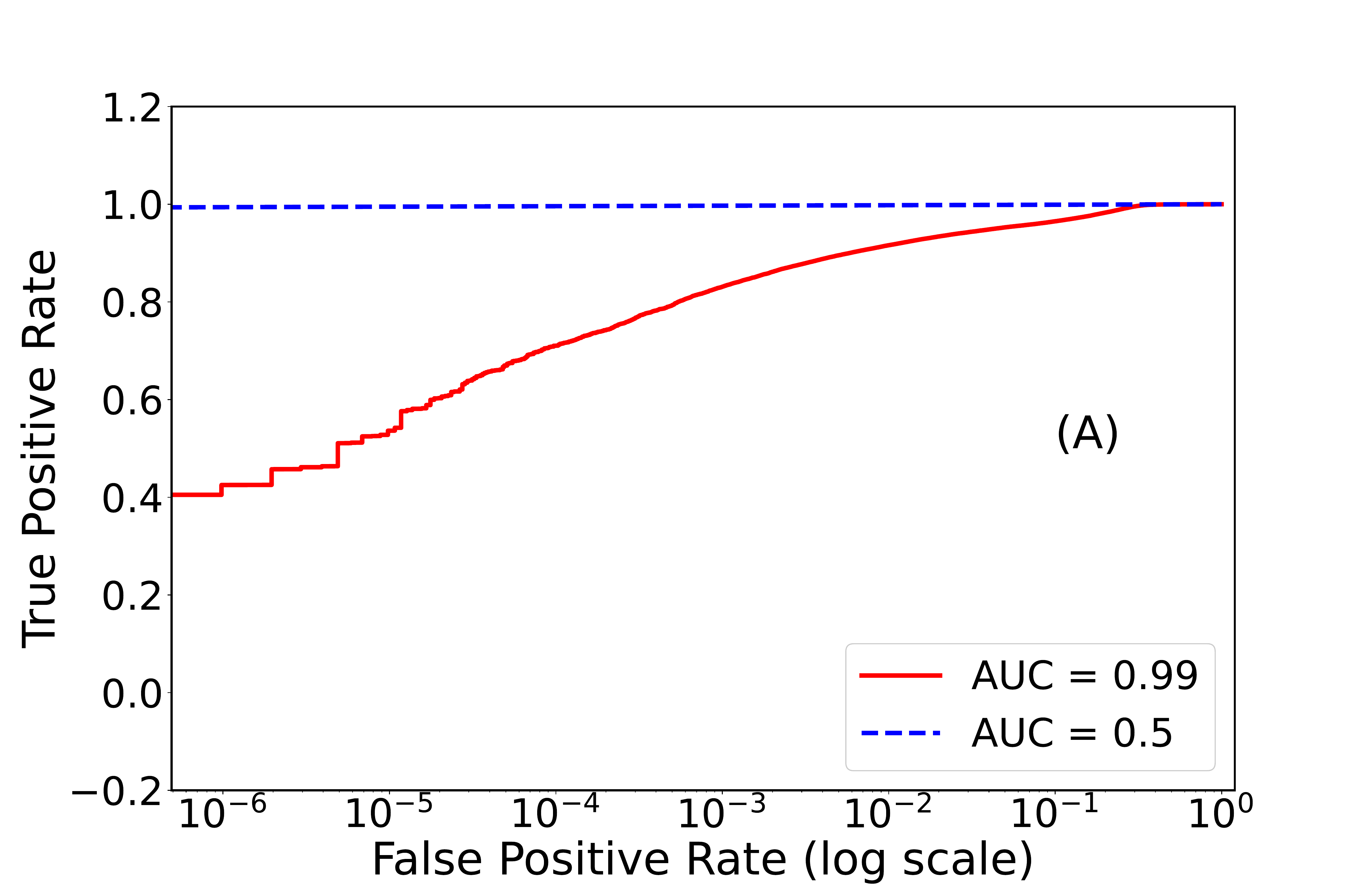}
    \includegraphics[width=0.32\textwidth]{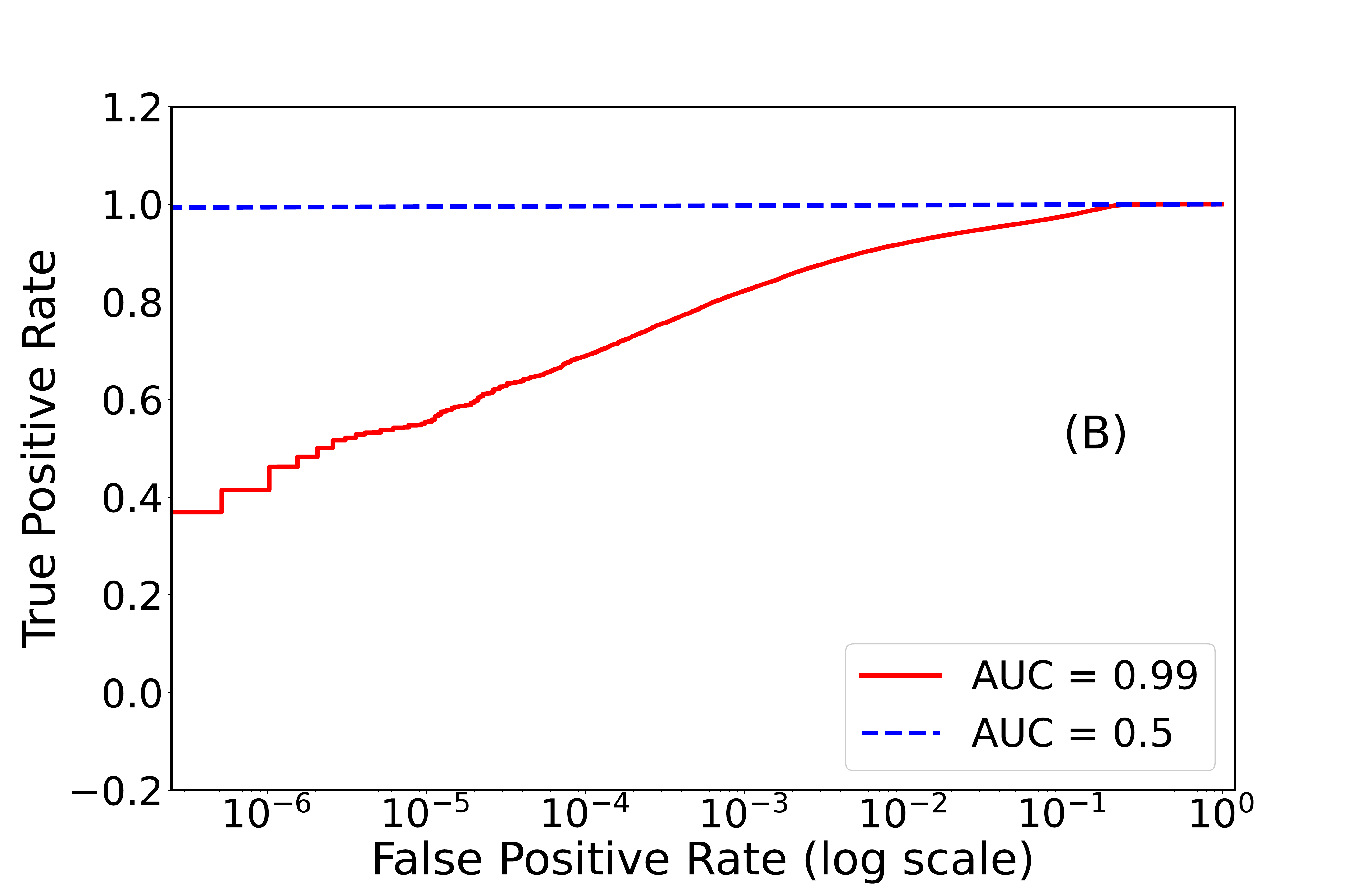}
    \includegraphics[width=0.32\textwidth]{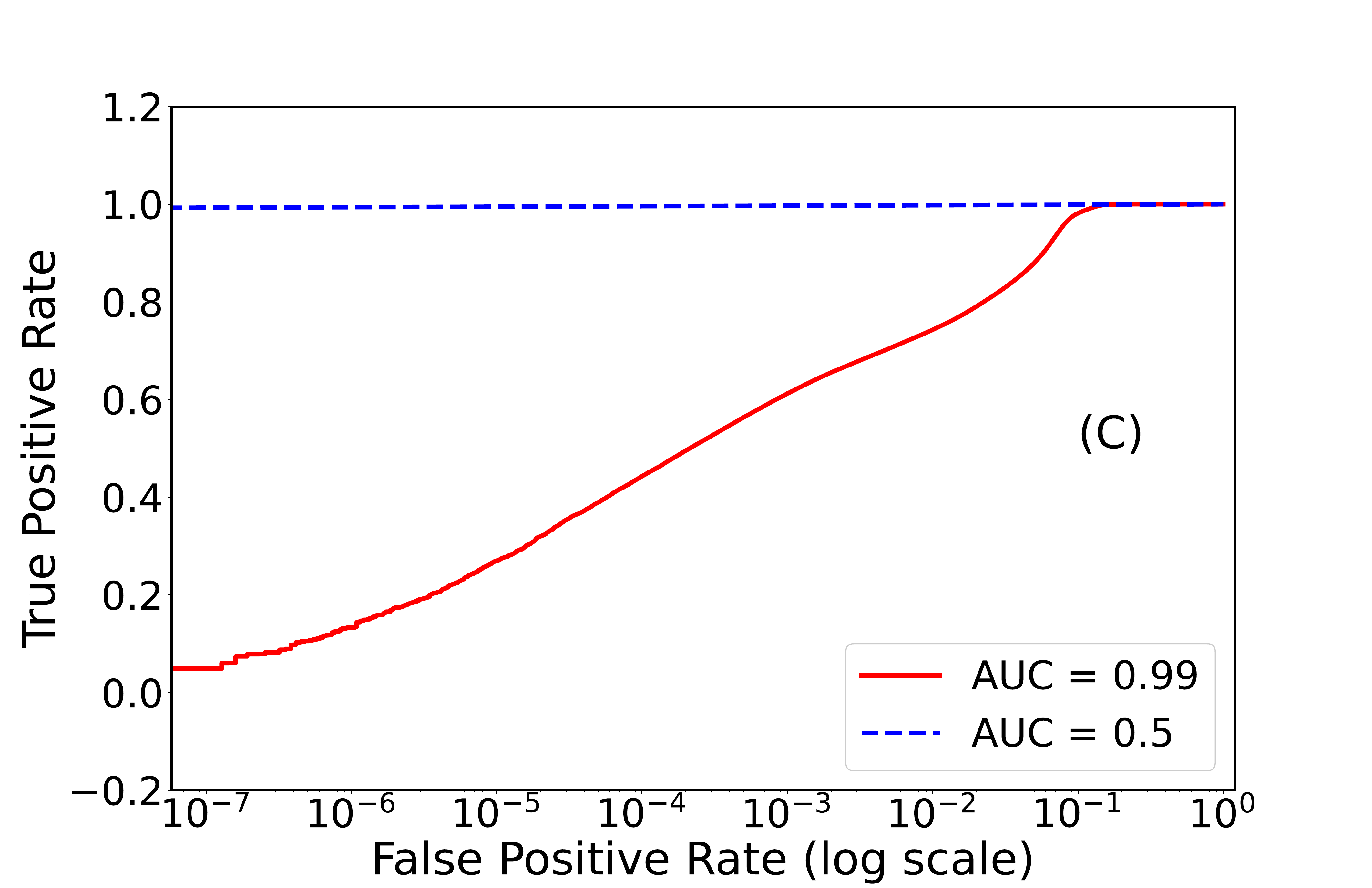}
    \caption{(color online) ROC Curve for various scenarios of graph network model}
    \label{fig:ROC}
\end{figure}

The Fig.~\ref{fig:ROC} gives and AUC Score of 0.99. We infer that the model is performing very well at distinguishing between the positive and negative classes ( primary particle track and other tracks). Moreover, We infer that model is robust towards mishits, double-hits because we  are using realistic simulation data which may account for mis-hits, double hits and other various challenges giving noise in the tracks.

\section{Conclusion}
The pipeline demonstrated the development of a novel approach for muon track reconstruction based on geometric deep learning. The method is based on Graph Neural Networks which manifest a significant improvement for solving track reconstruction tasks as compared to previous approaches by learning on a graph representation of hit data. The proposed method of cosmic ray muon tracking by using GNNs has performed with an accuracy of $\sim$ 97\% in finding true tracks. We believe that the GNN approach is a promising deep learning solution for addressing and solving the problems in tracking at high luminosity experiments like HL-LHC~\cite{hllhc} and can also become fruitful and applicable in the field of astroparticle experiments like IceCube~\cite{icecube}, GRAPES-3~\cite{grapes} experiments where muon tracking is considered essential.  
One requirement is to further explore and scale up this idea is to test and demonstrate the capability of these methods in a realistic tracking environment with full complexity into a complete tracking solution with robust, physics-driven graph construction method i.e. application to real experiment meta-data, pile-ups, irregular detector geometry, and various experimental constraints.   
\label{sec:conc}

\acknowledgments

The simulation and training works were carried-out in the computing facility in EHEP Lab at IISER Mohali.

\bibliographystyle{unsrt}
\bibliography{Tracking.bib}


\end{document}